\documentclass[aps,preprint,onecolumn,secnumarabic,nobalancelastpage,amsmath,amssymb,
nofootinbib]{revtex4}

\RequirePackage{fix-cm}
\usepackage{enumerate}

\usepackage{pst-plot}
\usepackage{pstricks,pstricks-add}
\usepackage[hang,nooneline,scriptsize]{subfigure}

\usepackage{graphicx}      
\usepackage{url}           
\usepackage{dcolumn}
\usepackage{bm}            
\usepackage{mathrsfs}
\usepackage{epstopdf}
\usepackage{color}

\begin{document}

\preprint{APS/123-QED}
\title{Investigation the geodesic motion of three dimensional rotating black holes}

\author{Sobhan Kazempour${}^1$}
\author{Saheb Soroushfar${}^{2}$}
\email{soroush@yu.ac.ir}
 \affiliation{${}^1$Department of Physics, University of Guilan, 41335-1914, Rasht, Iran \\
 ${}^2$ Faculty of Technology and Mining, Yasouj University, Choram 75761-59836, Iran \\}
\date{\today}

\begin{abstract}
We study the geodesic equations in the space-time of neutral Brans-Dicke Dilaton black hole in three dimensions, BTZ black holes and the 2+1 black hole. We use the process of separation of the Hamilton-Jacobi equation to obtain the constants of motion. The whole analytical solution of the geodesic equations in the space-times of the intended black holes are shown completely. Moreover, the geodesic equations are solved in terms of Weierstrass elliptic functions. Furthermore, with use of  the analytical solution and effective potential technique some trajectories around the black holes are classified. Meanwhile, by analytical solution, effective potential and considering the zeroes of underlying polynomials, some possible orbits are plotted. Finally, we compare our results with Cruz {\it et. al.} \cite{Cruz:1994ar} and we indicate the benefits of the analytical method which is applied in this paper. 
\end{abstract}

\maketitle

\section{INTRODUCTION}

Schwarzschild introduced the first exact solution of the vacuum Einstein equations in four dimensions which describes a spherically symmetric black hole \cite{Schwarzschild:1916ae}. Black holes are exact solutions of several theories of gravity in four, lower or higher dimensions, however black holes in four dimensions in General Relativity emerge from complete gravitational collapse of massive objects, such as stars \cite{Oppenheimer:1939ue}. Meanwhile, studying black holes in lower than four dimensions helps to a better understanding and analysing of conceptual issues of thermodynamic features such as entropy and radiated flux in a black hole geometry \cite{Harvey:1993,Horowitz:1992,Lemos:1995cp,Ashtekar:2002qc}.

As one part of our study in this research, Brans-Dicke theory which describes gravitation in terms of a scalar field as its new ingredients and accommodates both Mach’s principle and Dirac’s large-number hypothesis, is the simplest modification of General Relativity in its mathematical structure \cite{Brans:1961sx}. However, the singularity problem remains in Brans-Dicke theory, all the available observational and experimental tests being passed \cite{Clifford:1993}. For turning the 3-dimensional gravity much more similar to the realistic 4-dimensional, the dilaton gravity was introduced by coupling a scalar field to Einstein gravity with introducing local degrees of freedom. The local dynamical degree of freedom to the theory is provided by the scalar field and it appears naturally in string theory. Moreover, one of these theories was studied by Dias and Lemos is an Einstein-dilaton gravity of the Brans-Dick type in a $\Lambda <0$ background. This theory is specified by a Brans-Dicke parameter $\omega $. Here we use neutral Brans-Dicke Dilaton black hole for $\omega =-2$ which was prepared by Dias and Lemos \cite{CamposDias:2003tv}.

As another part, Banados, Teitelboim and Zanelli (BTZ) derived a three-dimensional black hole with a negative cosmological constant \cite{Banados:1992wn,Ross:1992ba},  which in its more complicated extension added by rotation and electrical charge \cite{Clement:1993kc,Martinez:1999qi}, is useful to study of black holes in quantum scales \cite{Carlip:1995qv,Strominger:1996sh}. Three dimensional BTZ black hole and a fewer dimensional black hole in General Relativity have common characteristics such as gravitational collapse of matter \cite{Ross:1992ba}, thermodynamic properties such as temperature and entropy \cite{Brown:1994gs} and their description in a three dimensional perfect fluid star \cite{Cruz:1994ar}. In addition, BTZ black hole can be transformed into a four-dimensional black string in General Relativity \cite{Ashtekar:2002qc,Sa:1996ty}. The BTZ black hole is also a solution to string theory in three dimensions \cite{Horowitz:1993jc} in which a great kind of black hole solutions such as black strings and black membranes, have been created, as same as being a solution of low energy string theory with a non-vanishing anti-symmetric tensor \cite{Horowitz:1993jc,Kaloper:1993kj}. On the other hand, the 2+1 dimensional black hole which specifies by mass, angular momentum and charge and defines by flux integrals at infinity, is quite similar to its 3+1 counterpart.

In order to study the observable effects such as light deflection, gravitational time-delay, perihelion shift and the Lense-Thirring effect and also to study the gravitational fields of black holes with studying the motion of test particles and light rays, there are two approaches to solve the problems: the numerical method and the analytical method. 
Due to the higher acuuracy of analytical method, we use it in this article and in order to simplify expressions we use the Weierstrassian elliptic function.
In history of analytical solution, Hagihara used elliptic functions to solve the geodesic equation in Schwarzschild spacetime analytically \cite{Hagihara:1931}, but then the interest faded out, however, the theory of Weierstrass elliptic functions is still a standard part of function theory. 
Recently, there has been a trend toward elliptic and hyperelliptic functions in order to solve geodesic equations analytically,  \cite{Soroushfar:2015wqa,Fujita:2009bp,Grunau:2012ri,Kraniotis:2004cz,Kraniotis:2003ig,Soroushfar:2015dfz,Soroushfar:2016yea,Hoseini:2016nzw,Soroushfar:2016esy,Kazempour:2016dco}.
Grunau and Kagramanova discussed analytical solutions of the electrically and magnetically charged test particles \cite{Grunau:2010gd}. Chandrasekhar made the study of time-like geodesics in a Schwarzschild space-time \cite{Chandrasekhar:1983}. Maki and Shiraishi discussed motion of test particles around a charged dilatonic black hole \cite{Maki:1992up}. Bhadra studies  about gravitational lensing by a charged black hole in string theory and Fernando studied about the geodesics of the 2+1 dimensional string black hole \cite{Bhadra:2003zs,Fernando:2003gg}.

In this paper, we study the equations of motion and we show the analytical solutions for rotating space-times namely neutral Brans-Dicke Dilaton black hole, BTZ black hole and the 2+1 black hole. For a better understanding of 2+1 black hole solutions, we have compared our results with which obtained by Cruz {\it et. al.}, \cite{Cruz:1994ar}.
 
Our paper is organized as follows: in Section \ref{Sec2}, we introduce the metrics. In Section \ref{Sec3}, we use the process of separation of the Hamilton-Jacobi equations to obtain the constants of motion, in section \ref{Asge} we present the analytical solution of the equations of motion, and in section \ref{Sec5} we analyse the possible orbit types and classification orbits using effective potential techniques and underlying polynomials. Furthermore, in section \ref{Sec6} we plot the possible orbits of test particle and light rays in these three space-times. The equations of motion are of elliptic type and the solutions are given in terms of the Weierstrass functions. 
\clearpage
\section{The spacetimes}\label{Sec2}

\subsection{Neutral Brans-Dicke Dilaton black hole for $\omega =-2$} 
The Brans-Dicke theory of gravitation is considered a viable alternative to General Relativity. It introduces an additional long-range scalar field $\Phi$ besides the metric tensor $g_{\mu\nu}$ of spacetime. The gravitational constant $G$ is not presumed to be constant, it is proportional to the inverse of the scalar field. In addition, the field equation of Brans-Dicke theory include a parameter $\omega$ which called the Brans-Dicke coupling constant. Light deflection and precession of perihelia of planets orbiting the sun can be predicted by Brans-Dicke theory \cite{Brans:1961sx}.
This black hole is derived by considering the Brans-Dicke action in three dimensions
\begin{align}\label{1}
S=\frac{1}{2\pi}\int d^{3}x\sqrt{-g}e^{-2\Phi}[R-4\omega(\partial\Phi)^{2}+\Lambda],
\end{align}
Where $g$ is the determinant of the three-dimensional metric, $R$ is the curvature scalar, $\Phi$ is a scalar field called dilaton $\Lambda<0$ is the cosmological constant, $\omega$ is the three-dimensional Brans-Dicke parameter. We are considering is given by
\begin{equation}\label{2}
{ds^{2}}_{1}=-(r^{2}-\frac{J^{2}}{M}r)dt^{2}-2Jr d\varphi dt+\dfrac{dr^{2}}{r^{2}-M(\frac{J^{2}}{M^{2}}-1)r}+(r^{2}+Mr)d\varphi^{2}, 
\end{equation}
where $\omega =-2$, $|J|>M$, in which $M$ and $J$ are, respectively, the mass and the angular momentum \cite{CamposDias:2003tv}. Index $1$ refers to the neutral Brans-Dicke Dilaton black hole in whole paper. There is a horizon defined by $g^{rr}=0$ and are given by
\begin{equation}
r=\frac{J^{2}-M^{2}}{M}.
\end{equation}
\subsection{3D BTZ black hole} 
The BTZ black hole is similar to the kerr black hole which characterized by mass $M$, angular momentum $J$ and two horizons which called event horizon and inner horizon \cite{Banados:1992gq}. The BTZ black hole can be expressed through isometry group of the AdS spacetime. In order to remove the usual problems with closed timelike curves, the origin must be a topological singularity \cite{CamposDias:2003tv}.
This black hole is derived by considering the gravitational Einstein-Hilbert action in three dimensions 
\begin{equation}
S=\frac{1}{16\pi}\int d^{3}x\sqrt{-g}(R-2\Lambda),
\end{equation}
Where $g$ is the determinant of the metric, $R$ is the Ricci scalar and $\Lambda$ is the cosmological constant. The 3D BTZ black hole appears as a solution of Einstein's gravity with a negative cosmological term, $\alpha^{2}\equiv-\Lambda>0,$ and zero 3D energy-momentum tensor, $T_{ab}^{(3)}=0$. The metric is given by \cite{Lemos:1995cp},
\begin{align}\label{S2}
{ds^{2}}_{2}=-(\alpha^{2}r^{2}-8M)dt^{2}-8Jd\varphi dt+\dfrac{dr^{2}}{\alpha^{2}r^{2}-8M+\frac{16J^{2}}{r^{2}}}+r^{2}d\varphi^{2},
\end{align}
where $M$ and $J$ are the mass and angular momentum of the black hole.  Index $2$ refers to the 3D BTZ black hole in whole paper. Horizons which are defined by $g^{rr}=0$, obtain as \cite{Lemos:1995cp,Banados:1992wn}
\begin{equation}
r_{\pm}=\pm\dfrac{2\sqrt{M \pm \sqrt{-J^{2}\alpha^{2}+M^{2}}}}{\alpha},
\end{equation}
where $M>0$ and $|J|\leq\frac{M}{\alpha}$. In the following, the 2+1 black hole which is studied by Cruz {\it et. al.} \cite{Cruz:1994ar} has been introduced:
\begin{align}\label{S3}
{ds^{2}}_{3}=-N^{2}dt^{2}+N^{-2}dr^{2}+r^{2}(N^{\phi}dt+d\phi)^{2}, \nonumber \\ N^{2}=-M+\frac{r^{2}}{l^{2}}+\frac{J^{2}}{4r^{2}}, \qquad N^{\phi}=-\frac{J}{2r^{2}}, \qquad l=(-\Lambda)^{-\frac{1}{2}},
\end{align}
where $ -\infty <t<\infty$, $0<r<\infty$ and $0\leqslant \phi <2\pi$. Index 3 refers to the 2+1 black hole in whole paper. Horizons of this spacetime which are obtained as solutions of $g^{rr}=0$ illustrate below
\begin{equation}
r_{\pm}=\frac{1}{2}\sqrt{2l^{2}M\pm 2l\sqrt{l^{2}M^{2}-J^{2}}}=lM^{\frac{1}{2}}\bigg[\frac{1\pm\sqrt{1-(\frac{J}{Ml})^{2}}}{2}\bigg]^{\frac{1}{2}}.
\end{equation}
\section{The Geodesic Equations}\label{Sec3}

In this section, using the Hamilton-Jacobi formalism, we derive the equations of motion for a neutral Brans-Dicke Dilaton black hole, Eq.(\ref{2}), BTZ black hole, Eq.(\ref{S2}) and the 2+1 black hole, Eq.(\ref{S3}) respectively, then we introduce effective potentials for the $r$ motion.

The Hamilton-Jacobi equation 
\begin{equation}\label{Hamilton0}
\dfrac{\partial S}{\partial\tau}+\frac{1}{2}\ g^{ij}\dfrac{\partial S}{\partial X^{i}}\dfrac{\partial S}{\partial X^{j}}=0,
\end{equation}
can be solved with an ansatz for the action
\begin{equation}
\label{S0}
S=\frac{1}{2}\varepsilon \tau - Et+L\varphi + S_{r} (r),
\end{equation}
in which the constants of motion are the energy $E$ and the angular momentum $L$ which are related to the generalized momenta $P_{t}$ and $P_{\varphi}$:
\begin{align}\label{P}
P_{t}=g_{tt}\dot{t}+g_{t\varphi}\dot{\varphi}=-E, \qquad\qquad P_{\varphi}=g_{\varphi\varphi}\dot{\varphi}+g_{t\varphi}\dot{t}=L,
\end{align}
and $\tau$ is an affine parameter along the geodesic, the parameter $\varepsilon$ is equal to 1 for particles and equal to 0 for light.\\
Plugining Eq.(\ref{S0}) into Eq.(\ref{Hamilton0}) for three metrics, Eq.(\ref{2}), Eq.(\ref{S2}) and Eq.(\ref{S3}), obtain Eq.(\ref{ds1}), Eq.(\ref{ds2}) and Eq.(\ref{ds3}), respectively
\begin{align}\label{ds1}
\varepsilon +\dfrac{M(r+M)E^{2}}{r^{2}(J^{2}-Mr-M^{2})}-\dfrac{r(J^{2}-Mr-M^{2})}{M}\left(\frac{dS}{dr}\right)_{1}^{2}+ \dfrac{(J^{2}-Mr)L^{2}}{r^{2}(J^{2}-Mr-M^{2})}\nonumber\\ -\dfrac{2JMEL}{r^{2}(J^{2}-Mr-M^{2})}=0.\qquad\qquad\qquad
\end{align}
\begin{align}\label{ds2}
\varepsilon -\dfrac{r^{2}E^{2}}{(\alpha^{2}r^{4}-8Mr^{2}+16J^{2})}+\dfrac{(\alpha^{2}r^{4}-8Mr^{2}+16J^{2})}{r^{2}}\left(\frac{dS}{dr}\right)_{2}^{2}+\dfrac{(\alpha^{2}r^{2}-8M)L^{2}}{(\alpha^{2}r^{4}-8Mr^{2}+16J^{2})}\nonumber\\ +\dfrac{8JEL}{(\alpha^{2}r^{4}-8Mr^{2}+16J^{2})}=0.\qquad\qquad\qquad
\end{align}
\begin{align}\label{ds3}
\varepsilon - \dfrac{4 l^{2}r^{2}E^{2}}{(-4Ml^{2}r^{2}+J^{2}l^{2}+4r^{4})}+\dfrac{(-4Ml^{2}r^{2}+J^{2}l^{2}+4r^{4})}{4l^{2}r^{2}}\left(\frac{dS}{dr}\right)_{3}^{2}-\dfrac{4(Ml^{2}-r^{2})L^{2}}{(-4Ml^{2}r^{2}+J^{2}l^{2}+4r^{4})}\nonumber\\+\dfrac{4Jl^{2}EL}{(-4Ml^{2}r^{2}+J^{2}l^{2}+4r^{4})}=0. \qquad\qquad\qquad
\end{align}
With the separation ansatz (\ref{S0}) we derive the equations of motion 
\begin{align}
\left(\frac{dr}{d\tau}\right)_{1}^{2}=\dfrac{(J^{2}-Mr-M^{2})\varepsilon r}{M}+\dfrac{(r+M)E^{2}}{r}+\dfrac{L^{2}(J^{2}-Mr)}{rM}-\dfrac{2JEL}{r}.
\end{align}
\begin{align}
\left(\frac{dr}{d\tau}\right)_{2}^{2}=-\dfrac{\varepsilon}{r^{2}}(\alpha^{2}r^{4}-8Mr^{2}+16J^{2})+E^{2}-\dfrac{L^{2}}{r^{2}}(\alpha^{2}r^{2}-8M)-\frac{8JEL}{r^{2}}.
\end{align}
\begin{align}
\left(\frac{dr}{d\tau}\right)_{3}^{2}=\dfrac{-\varepsilon (-4M l^{2}r^{2}+J^{2}l^{2}+4r^{4})}{4 l^{2}r^{2}} +E^{2}+\dfrac{(M l^{2}-r^{2})L^{2}}{l^{2}r^{2}}- \dfrac{JEL}{r^{2}}.
\end{align}
Introducing a new parameter, the so-called Mino time \cite{Mino:2003yg}, given by $d\tau=rd\lambda$, equations of motion obtain as:
\begin{align}\label{171}
\left(\frac{dr}{d\lambda}\right)_{1}^{2}=-\varepsilon r^{4}+\dfrac{(J^{2}\varepsilon -M^{2}\varepsilon)}{M}r^{3}+(E^{2}-L^{2})r^{2}+(ME^{2}+\frac{L^{2}J^{2}}{M}-2JEL)r.
\end{align}
\begin{align}\label{181}
\left(\frac{dr}{d\lambda}\right)_{2}^{2}=-\varepsilon \alpha^{2}r^{4}+(8M\varepsilon +E^{2} L^{2}\alpha^{2})r^{2}+(-16J^{2}\varepsilon -8JEL+8L^{2}M).
\end{align}
\begin{align}\label{191}
\left(\frac{dr}{d\lambda}\right)_{3}^{2}=\frac{-\varepsilon}{l^{2}} r^{4}+(\varepsilon M+E^{2}-\frac{L^{2}}{l^{2}})r^{2}+(ML^{2}-JEL-\frac{\varepsilon J^{2}}{4}).
\end{align}
Moreover, the possible orbits can be imagined by effective potentials, which can be solved explicitly with  corresponding $\left(\frac{dr}{d\lambda}\right)_{1}^{2}=0$, $\left(\frac{dr}{d\lambda}\right)_{2}^{2}=0$ and $\left(\frac{dr}{d\lambda}\right)_{3}^{2}=0$, such that $V_{1,2,3_{eff}}=E$. As a result, Eqs. (\ref{V11},\ref{V22},\ref{V33}) can be obtained:
\begin{align}\label{V11}
V_{1_{eff}}=\dfrac{LJM\pm \sqrt{-Mr(J^{2}-Mr-M^{2})(\varepsilon r^{2}+L^{2}+M\varepsilon r)}}{M(r+M)}.
\end{align}
\begin{align}\label{V22}
V_{2_{eff}}=\dfrac{4JL\pm \sqrt{(\alpha^{2}r^{4}-8Mr^{2}+16J^{2})(\varepsilon r^{2}+L^{2})}}{r^{2}}.
\end{align}
\begin{align}\label{V33}
V_{3_{eff}}=\dfrac{JLl\pm \sqrt{4\varepsilon r^{6}+(4L^{2}-4M\varepsilon l^{2})r^{4}+(\varepsilon J^{2}l^{2}-4L^{2}Ml^{2})r^{2}+J^{2}L^{2}l^{2}}}{2r^{2}l}.
\end{align}
\newpage
\section{Analytical solution of geodesic equations}\label{Asge}
In this section, we present the analytical solution of the equations of motion for neutral Brans-Dicke Dilaton, 3D BTZ black holes and the 2+1 black hole. 
\subsection{Neutral Brans-Dicke Dilaton black hole}
We introduce a new variable $u=\frac{1}{r}$ and obtain from Eq.(\ref{171}):
\begin{align}\label{dr/du1}
\left(\frac{du}{d\lambda}\right)_{1}^{2}=(\frac{J^{2}L^{2}}{M}-2JEL+ME^{2})u^{3}+(E^{2}-L^{2})u^{2}+(\frac{J^{2}\varepsilon -M^{2}\varepsilon}{M})u-\varepsilon=\sum_{i=0}^{3}b_{i}u^{i}.
\end{align}
a further substitution $u=\frac{1}{b_{3}}(4y-\frac{b_{2}}{3})$ transforms $P_{3}(u)$ into the Weierstrass form so that Eqs.(\ref{dr/du1}) turns into:
\begin{align}\label{dy1}
(\frac{dy}{d\lambda})^{2}=4y^{3}-g_{2}y-g_{3}=P_{3}(y),
\end{align}
where
\begin{align}\label{g2g3}
g_{2}=\dfrac{b_{2}^{2}}{12}-\dfrac{b_{1}b_{3}}{4}, \qquad g_{3}=\dfrac{b_{1}b_{2}b_{3}}{48}-\dfrac{b_{0}b_{3}^{2}}{16}-\dfrac{b_{2}^{3}}{216}.
\end{align}
are the Weierstrass invariants. The differential equation (\ref{dy1}) is of elliptic type and is solved by the Weierstrass function \cite{Abramowitz:1968,Hackmann:2008zz,Whittaker:1973}. 
\begin{align}\label{Y}
y(\lambda)=\wp(\lambda -\lambda_{in};g_{2},g_{3}),
\end{align}
where $\lambda_{in}=\lambda_{0}+\int_{y_{0}}^{\infty}\dfrac{dy}{\sqrt{4y^{3}-g_{2}y-g_{3}}}$ with $y_{0}=\frac{b_{3}}{4r_{0}}+\frac{b_{2}}{12}$, depending only on the initial values $\lambda_{0}$ and $r_{0}$. Then the solution of Eq.(\ref{171}) acquires the form
\begin{align}
r(\lambda)=\dfrac{b_{3}}{4\wp(\lambda -\lambda_{in};g_{2},g_{3})-\frac{b_{2}}{3}}.
\end{align}
\subsection{3D BTZ and the 2+1 black holes}
We have polynomials of degree four in the form $R(r)=\sum_{i=0}^{4}c_{i}r^{i}$, with only simple zeros, which for solving them in this way, we can apply up to two substitutions. The first substitution is $r=\frac{1}{u}+r_{R}$, where $r_{R}$ is a zero of $R$, transforms the problem to  
\begin{align}
\left(\frac{du}{d\lambda}\right)_{2}^{2}=R_{3}(u)=\sum_{j=0}^{3}c_{j}u^{j}, \qquad\qquad u(\lambda_{0})=u_{0},
\end{align}
with a polynomial $R_{3}$ of degree 3. Where
\begin{align}
c_{j}=\dfrac{1}{(4-j)!}\dfrac{d^{(4-j)}R}{dr^{(4-j)}}(r_{R}),
\end{align}
In which $c_{j}$, $(j=1,2,3)$ is an arbitrary constant for each metric which is related to the parameter of the relevant metric. A second substitution $u=\frac{1}{c_{3}}(4y-\frac{c_{2}}{3})$, changes $R_{3}(u)$, into the Weierstrass form exact as Eqs.(\ref{dy1})-(\ref{Y}).\\
Where $\lambda_{in}=\lambda_{0}+\int_{y_{0}}^{\infty}\dfrac{dy}{\sqrt{4y^{3}-g_{2}y-g_{3}}}$ with $y_{0}=\frac{c_{3}}{4r_{0}}+\frac{c_{2}}{12}$, depending only on the initial values $\lambda_{0}$ and $r_{0}$. Than the solution of Eqs.(\ref{181},\ref{191}) acquires the form
\begin{align}
r(\lambda)=\dfrac{c_{3}}{4\wp(\lambda -\lambda_{in};g_{2},g_{3})-\frac{c_{2}}{3}}+r_{R}.
\end{align}
\section{Classification of the geodesics}\label{Sec5}
With the analytical solution derived in section \ref{Asge} and effective potential equations also by considering the zeroes of underlying polynomials, we plot figures and show some possible orbits. 
\\
Before that, we introduce a list of all possible orbits: 
\begin{enumerate}
\item \textit{Terminating orbit} (TO) with ranges $r\in[0,\infty)$. The TO end in the singularity at $r=0$.
\item \textit{Terminating bound orbit} (TBO), $r$ starts in $(0,r_{1}]$ for $0<r_{1}<\infty$ and falls into the singularity at $r=0$.
\item \textit{Escape orbit} (EO) with range $[r_{1},\infty)$ with $0<r_{1}$ or $(-\infty,r^{'}_{1}]$ with $r^{'}_{1}<0$. These escape orbits do not cross $r=0$.
\item \textit{Bound orbits} (BO) with range $r\in(r_{1},r_{2})$ with $r_{1}<r_{2}$ and\\
(a) Either $r_{1},r_{2}>0$ or\\
(b) $r_{1},r_{2}<0$.
\item \textit{Crossover bound orbits} (CBO) with range $r\in (r_{1},r_{2})$ where $r_{1}<0$ and $r_{2}>0$.
\end{enumerate}
\begin{itemize}
\item Region (A): one zero. We have TBO only with particles moving from $r_{1}$ to $r=0$
\item Region (B): no real zeroes. Only TO are possible. 
\item Region (C, D): four zeroes. We have bound orbits, two zeroes are positive and two zeroes are negative, in C region both of them are inside the inner horizons and the others are out of  outer horizons, in region D all of them are between the inner horizons and the outer horizons and the turning point can coincide with the event horizons.
\item Region (E): two zeroes. We have escape orbits, a zero is positive and another is negative, all of them are inside the horizons.
\item Region (F): two zeroes. We have Crossover bound orbits, a zero is positive and another is negative, all of them are outside the horizons.
\end{itemize}
As it was mentioned before, the motion is essentially characterized by the number of real zeroes the polynomials possess. If polynomials possess no real zero, then the particle and light are coming from infinity and move directly to the singularity at $r=0$. Suchlike an orbit we call a terminating orbit. Moreover, one real positive zero means the particle starts at a finite coordinate distance and ends at $r=0$, which we call a terminating bound orbit. For two real zeroes we can have two cases: (i) the light ray moves on an escape orbit and do not cross $r=0$. (ii) we have crossover bound orbit and it do cross $r=0$ several times. Furthermore, four real zeroes means that we have two bound orbits like a planetary orbit.
Summary of possible orbit types can be found in the Tables~\ref{table1}, \ref{table2} and \ref{table3}.
\subsection{Neutral Brans-Dicke Dilaton black hole for $\omega =-2$}
The effective potential for particle and light and the possible orbits for neutral Brans-Dicke Dilaton black hole are shown below.
\begin{figure}
\centering
\includegraphics[width=7cm]{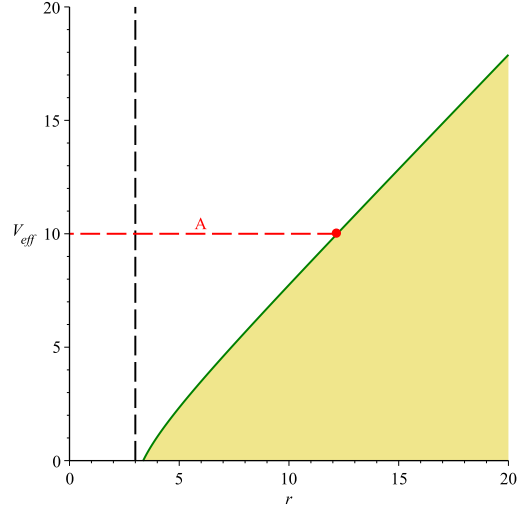}
\caption
{ Effective potential of neutral Brans-Dicke Dilaton black hole for particle with parameters $\varepsilon=1$, $J=-2$, $M=1$, $L=2.5$. The vertical black dashed line is horizon of black hole. Example energy of region A is given as red horizontal line and the red dot mark the zero of the polynomial R, which is the turning point of the orbit. The khaki area is a forbidden zone, where the geodesic motion is not possible}
\label{fig:EBP}
\end{figure}
\begin{figure}
\centering
\includegraphics[width=7cm]{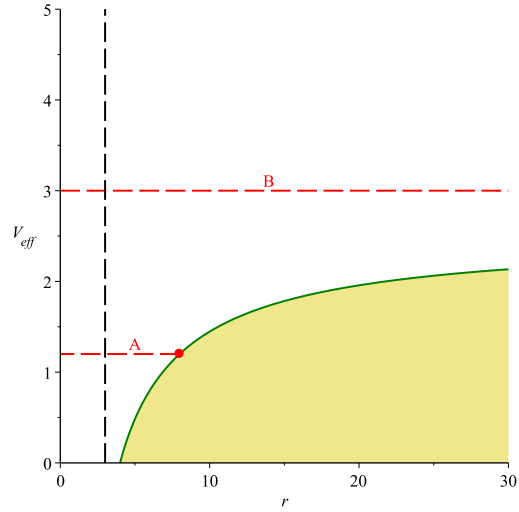}
\caption{ Effective potential of neutral Brans-Dicke Dilaton black hole for light ray with parameters $\varepsilon=0$, $J=-2$, $M=1$, $L=2.5$. The vertical black dashed line is horizon of black hole. Example energies of regions A and B are given as red horizontal lines and the red dots mark the zeros of the polynomial R, which are the turning points of the orbits. The khaki area is a forbidden zone, where the geodesic motion is not possible}
\label{fig:EBN}
\end{figure}
\begin{table}[!ht]
\begin{center}
\begin{tabular}{|l|l|c|l|}
\hline
Type & Zeroes & Range of $r$ &  Orbits \\
\hline\hline
A & 1 &
$|\textbf{---------}\lVert\textbf{------------}\bullet--------$
& TBO
\\  \hline
B & 0 &
$|\textbf{---------}\lVert\textbf{-----------------------------------}$
& TO
\\ \hline
\end{tabular}
\caption{Types of orbits of light and particles in the neutral Brans-Dicke Dilaton black hole. The lines represent the range of the orbits. The turning points are shown by thick dots. The horizon is indicated by a vertical double line.}
\label{table1}
\end{center}
\end{table}
\subsection{3D BTZ black hole} 
The effective potential and possible orbits for particle and light are indicated for 3D BTZ black hole.
\begin{table}[!ht]
\begin{center}
\begin{tabular}{|l|l|l|c|l|}
\hline
Type &-Zeroes &+Zeroes & Range of $r$ &  Orbits \\
\hline
C & 2 & 2 &
$--\bullet\textbf{------}\lVert\textbf{--------}\lVert\textbf{---}\bullet--|--\bullet\textbf{---}\lVert\textbf{--------}\lVert\textbf{------}\bullet--$
  & 2x BO
\\ \hline
D & 2 & 2 &
$----\lVert\bullet\textbf{-----}\bullet\lVert----|----\lVert\bullet\textbf{-----}\bullet\lVert----$
  & 2x BO

\\ \hline

E & 1 & 1 &
$\textbf{-----------}\lVert\textbf{--------}\lVert\textbf{---}\bullet--|--\bullet\textbf{---}\lVert\textbf{--------}\lVert\textbf{-----------}$
  & 2x EO
\\ \hline
\end{tabular}
\caption{Types of orbits of light and particles in 3D BTZ black hole. The lines represent the range of the orbits. The turning points are shown by thick dots. The horizons are indicated by a vertical double lines.}
\label{table2}
\end{center}
\end{table}
\begin{figure}[!ht]
\centering
\subfigure{\includegraphics[width=7.5cm]{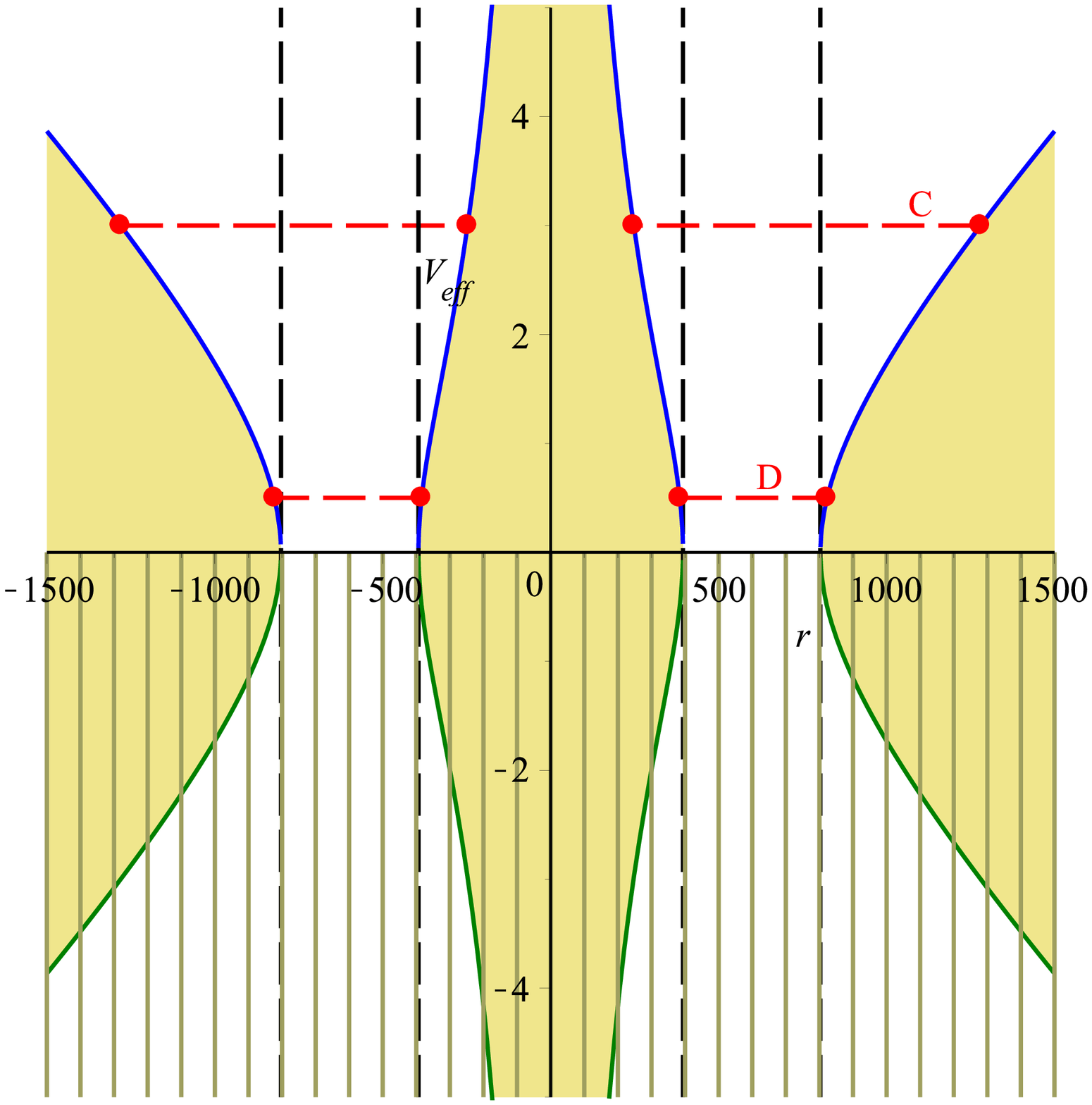}\label{C}}
\hspace*{1mm}
\subfigure{\includegraphics[width=7.5cm]{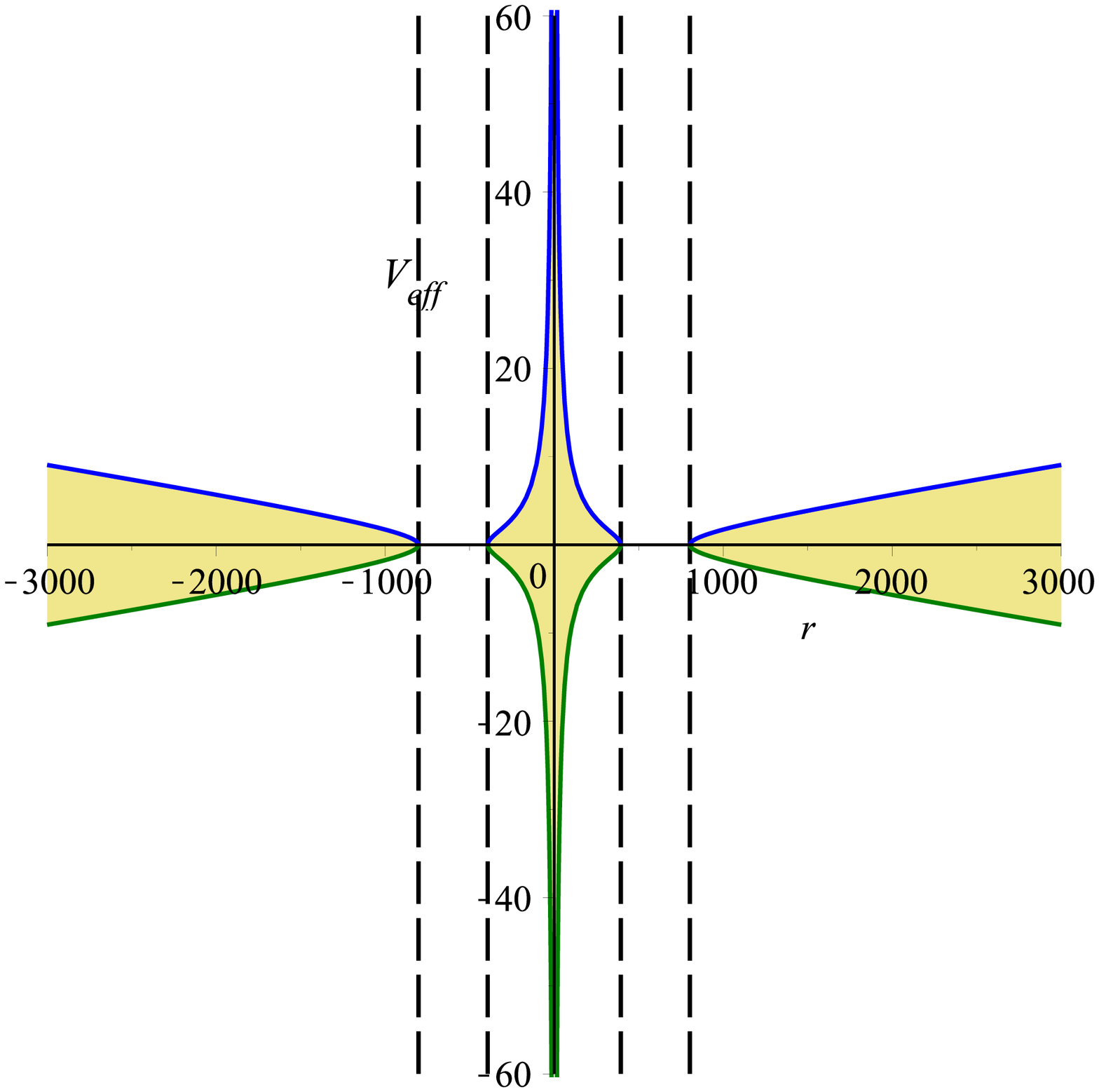}\label{D}}
\caption
{\subref{C} Effective potential of 3D BTZ black hole for particle with parameters $\varepsilon=1$, $J=250$, $M=1$, $L=0.5$, $\alpha^{2}=10^{-5}$. The vertical black dashed lines are horizons of black hole. Example energies of regions C and D are given as red horizontal lines and the red dots mark the zeros of the polynomial R, which are the turning points of the orbits. The khaki area is a forbidden zone, where the geodesic motion is not possible. \subref{D} overview.}
\label{EBTZP}
\end{figure}
\newpage
\begin{figure}
\centering
\subfigure{\includegraphics[width=7.5cm]{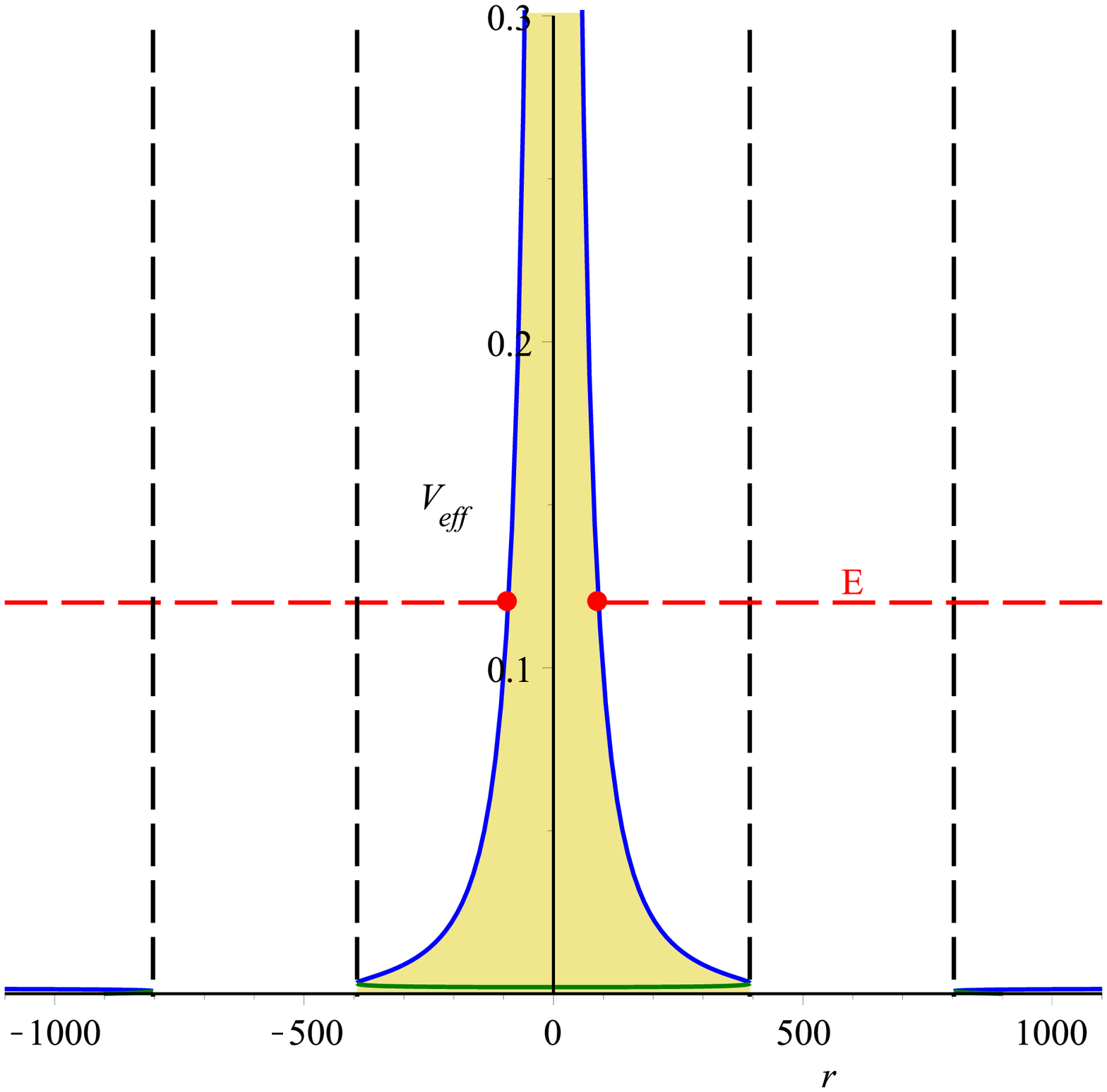}\label{E}}

\subfigure{\includegraphics[width=7.5cm]{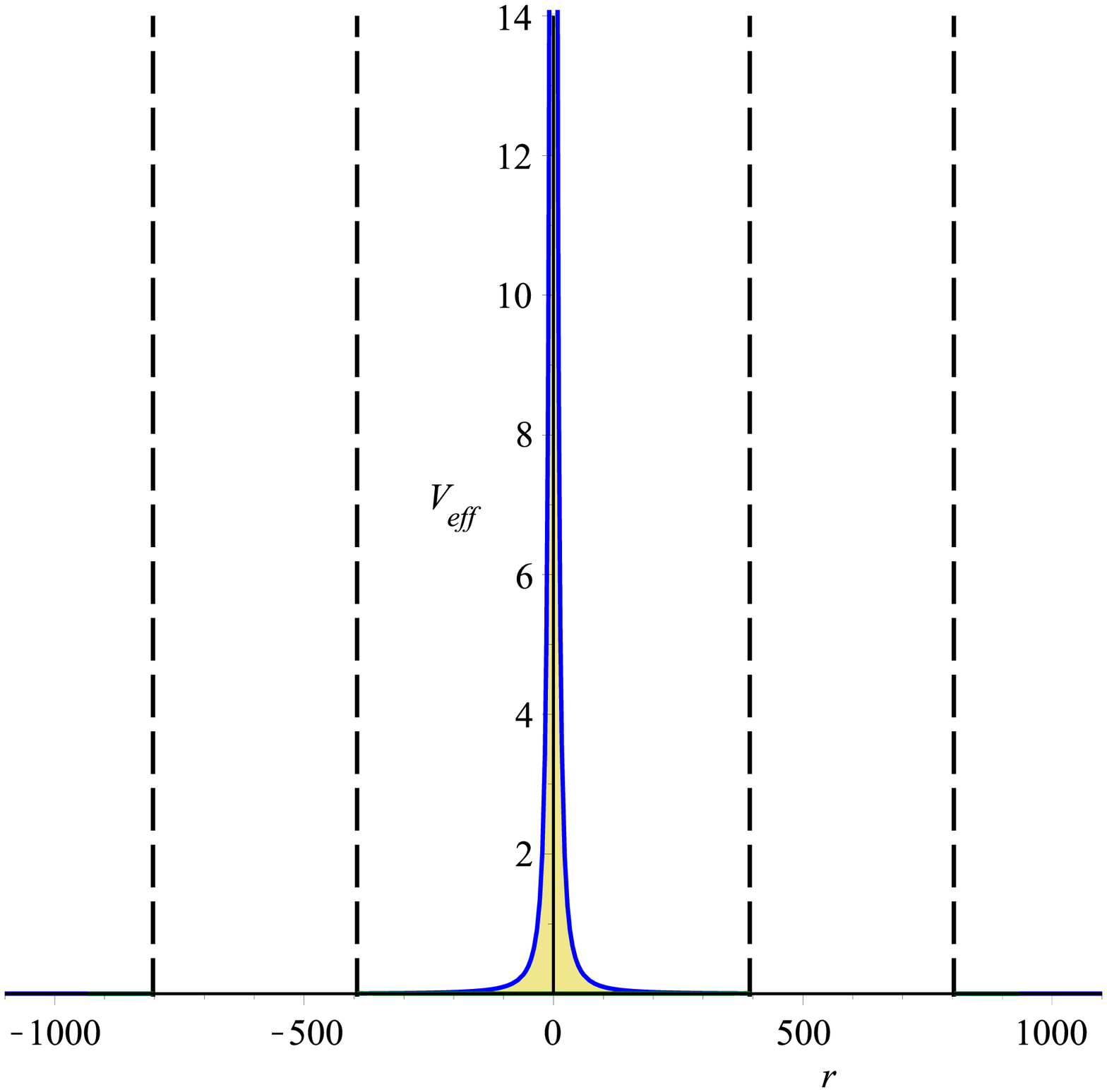}\label{E2}}
\caption
{\subref{E} Effective potential of 3D BTZ black hole for light ray with parameters $\varepsilon=0$, $J=250$, $M=1$, $L=0.5$, $\alpha^{2}=10^{-5}$. The vertical black dashed lines are horizons of black hole. Example energy of region E is given as red horizontal lines and the red dots mark the zeros of the polynomial R, which are the turning points of the orbits. The khaki area is a forbidden zone, where the geodesic motion is not possible. \subref{E2} overview.}
\label{EBTZN}
\end{figure}

\newpage
\subsection{The 2+1 black hole}\label{Sec5.3}
\begin{figure}[!ht]
\centering
\subfigure{\includegraphics[width=7.5cm]{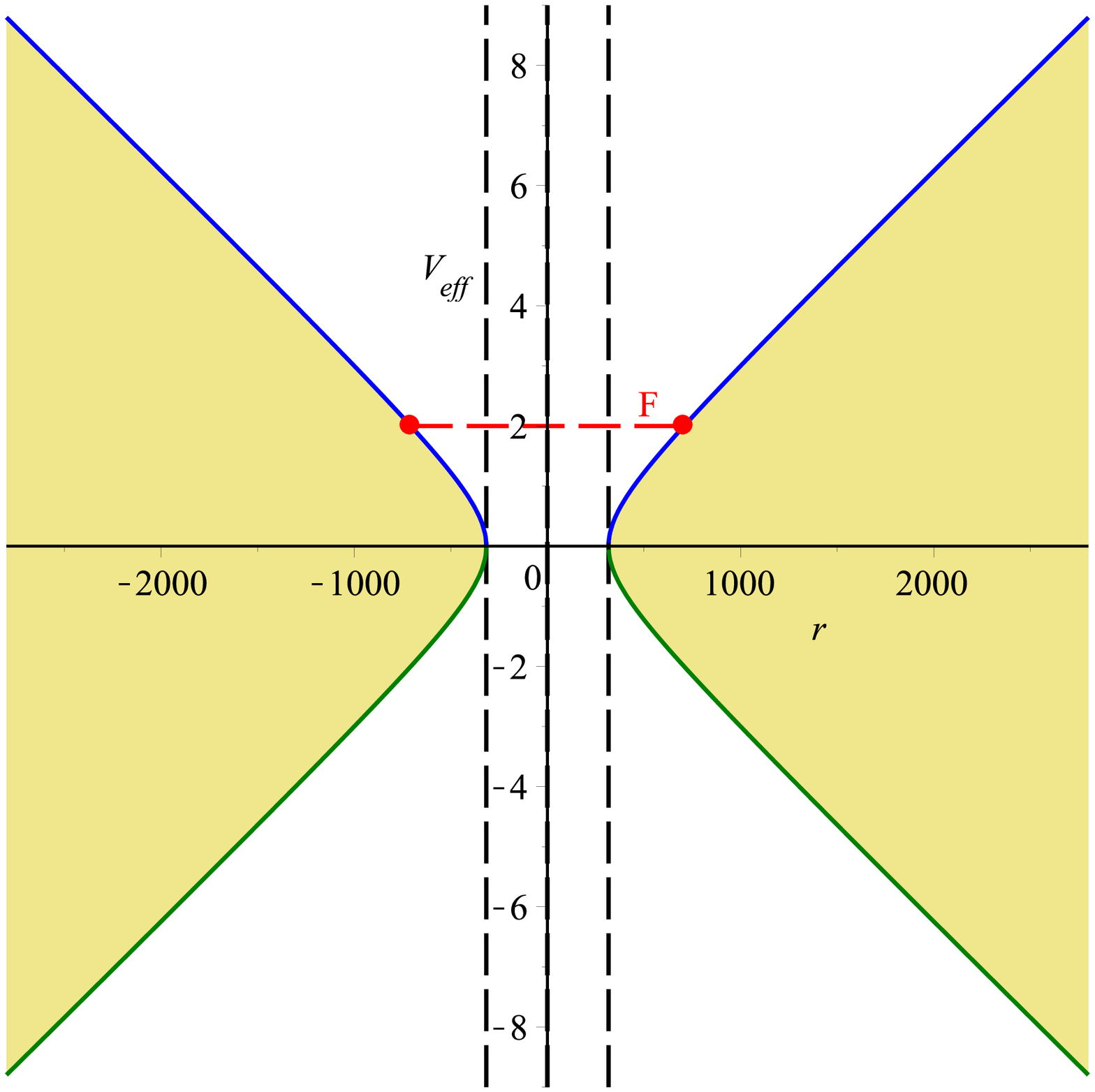}\label{MP}}
\hspace*{1mm}
\subfigure{\includegraphics[width=7.5cm]{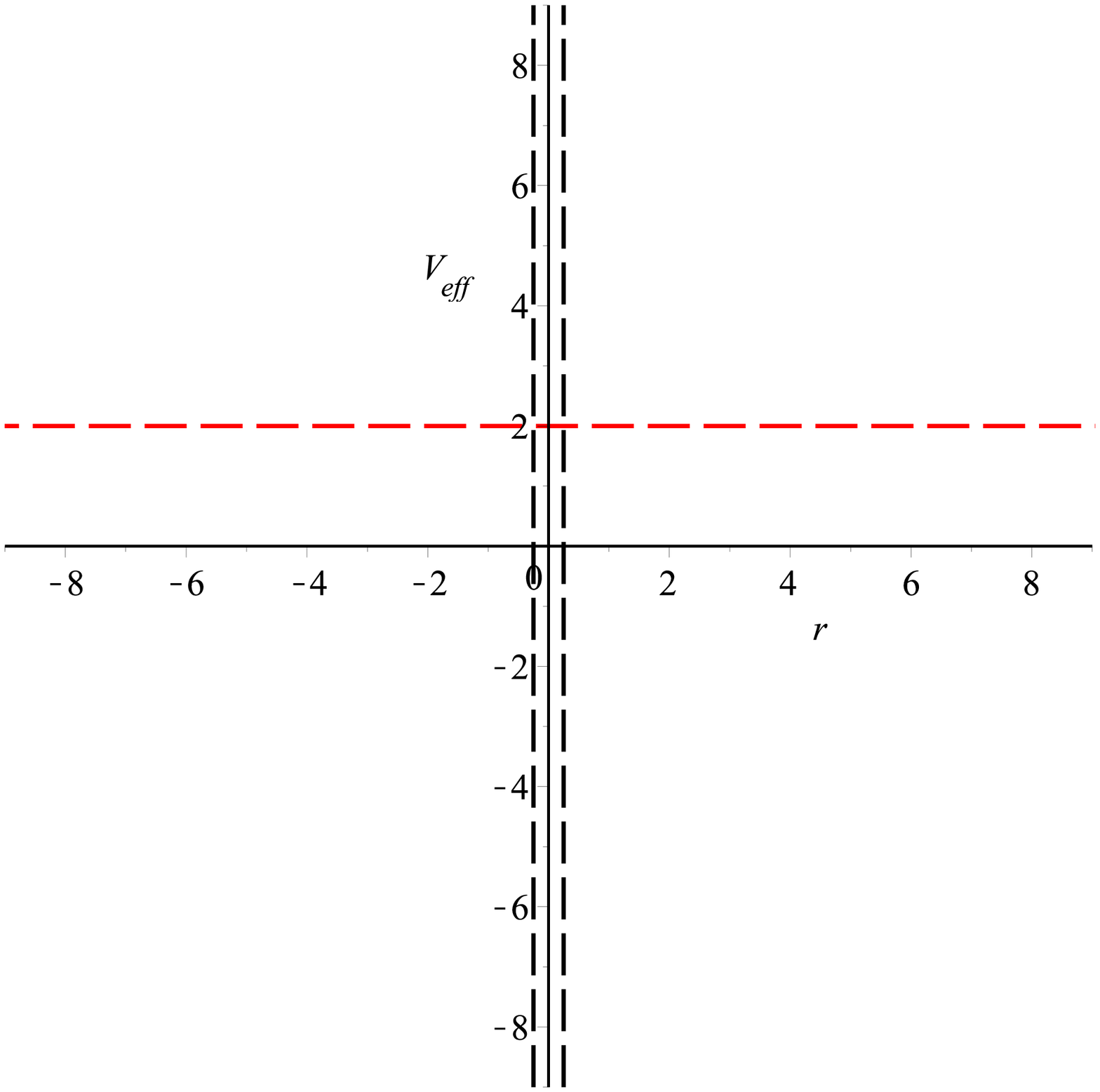}\label{MP1}}
\caption
{\subref{MP} Effective potential of the 2+1 black hole for particle with parameters $\varepsilon=1$, $J=0.5$, $M=1$, $L=2$, $E=2$, $l=\frac{1}{\sqrt{10^{-5}}}$. The vertical black dashed lines are horizons of black hole. Example energy of region E is given as red horizontal lines and the red dots mark the zeros of the polynomial R, which are the turning points of the orbits. The khaki area is a forbidden zone, where the geodesic motion is not possible. \subref{MP1}Close up view}
\label{EPCAM}
\end{figure}

In this part, the effective potential for the 2+1 black hole which was calculated in this paper, is plotted with the same amount of parameters which was used in Cruze and {\it et. al.} exactly. Here we point out that there is a characteristic difference between this effective potential and the effective potential which was plotted in Cruz and {\it et. al.}. The most striking feature is that there is crossover bound orbit and there are two zeroes as we show in figure \ref{EPCAM} and Table \ref{table3}. However, in the Cruz and {\it et. al.} there was terminating bound orbit and a zero \cite{Cruz:1994ar}. We show that there are two real zeroes and crossover bound orbit for the same parameters for this black hole. 
\begin{table}[!ht]
\begin{center}
\begin{tabular}{|l|l|l|c|l|}
\hline
Type & +Zeroes &-Zeroes &Range of $r$ &  Orbits \\
\hline
F & 1 &1&
$-----\bullet\textbf{-----}\lVert\textbf{-----------}\lVert|\lVert\textbf{-----------}\lVert\textbf{-----}\bullet-----$
  & CBO
\\ \hline
\end{tabular}
\caption{Type of orbit of particles in the 2+1 black hole. The lines represent the range of the orbits. The turning points are shown by thick dots. The horizon is indicated by a vertical double lines.}
\label{table3}
\end{center}
\end{table}
\newpage
\section{Orbits}\label{Sec6}
\begin{figure}[!ht]
\centering
\includegraphics[width=7.5cm]{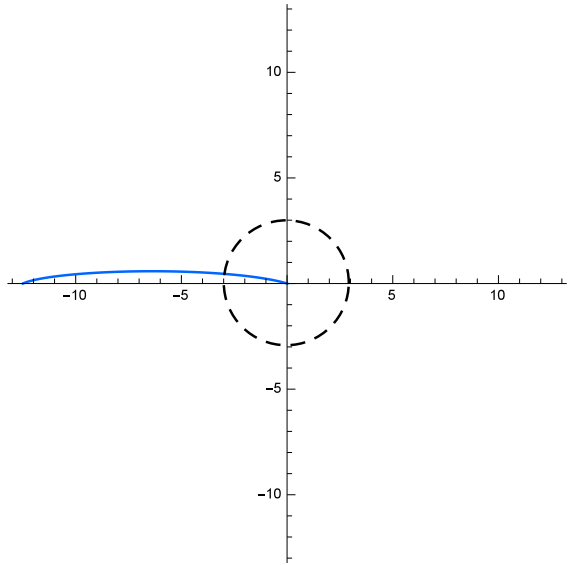}
\caption
{$E=10.5$, $L=2.5$, Terminating bound orbit in region (A) for particle.}
\label{O1}
\end{figure}

\begin{figure}[!ht]
\centering
\includegraphics[width=7.5cm]{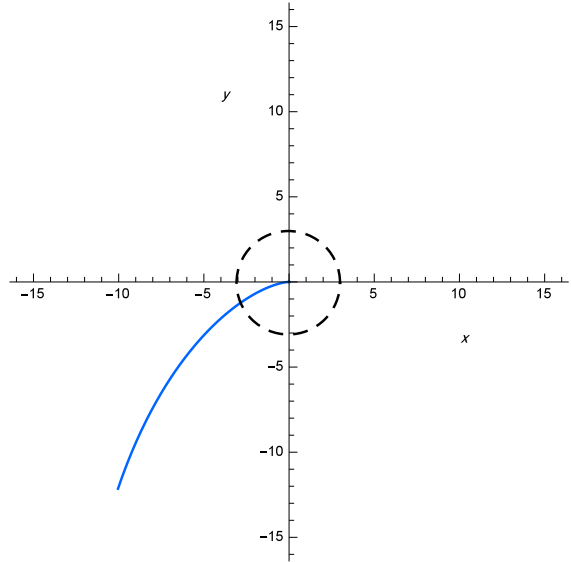}
	
\caption
{$E=3$, $L=2.5$, Terminating orbit in region (B) for light ray.}
\label{O2}
\end{figure}

\begin{figure}[!ht]
\centering
\includegraphics[width=7.5cm]{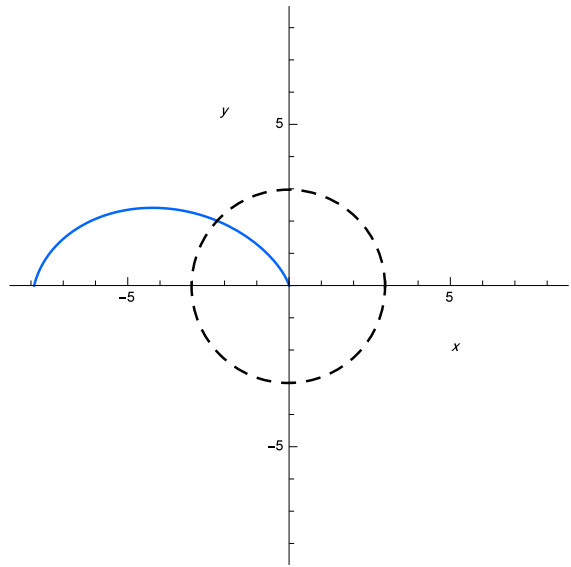}
	
\caption
{$E=1.5$, $L=2.5$, Terminating bound orbit in region (A) for light ray.}
\label{O3}
\end{figure}

The achived analytical results conduct us to some sets of orbits for light and test particles in the neutral Brans-Dicke Dilaton black hole, BTZ black hole and the 2+1 black hole. Depending on the parameters $\varepsilon$, $J$, $M$, $\alpha$, $L$ and $E$, terminating orbit, terminating bound orbit, escape orbit, bound orbits and crossover bound orbit are possible. Terminating bound orbits can be seen in figures \ref{O1} and \ref{O3}. Moreover, terminating orbit can be seen in figure \ref{O2}. Furthermore, bound orbits are showed in figures \ref{O4} and \ref{O5} respectively and figure \ref{EOBTZ} shows an escape orbit also crossover bound orbit can be seen in figure \ref{O8}. It is worth pointing out that there are not any bound orbit in neutral Brans-Dicke Dilaton black hole, while there are bound orbits in 3D BTZ black hole.
\subsection{Neutral Brans-Dicke Dilaton black hole for $\omega =-2$ orbits}

\clearpage

\subsection{3D BTZ black hole orbits }
\begin{figure}[!ht]
\centering
\includegraphics[width=7.5cm]{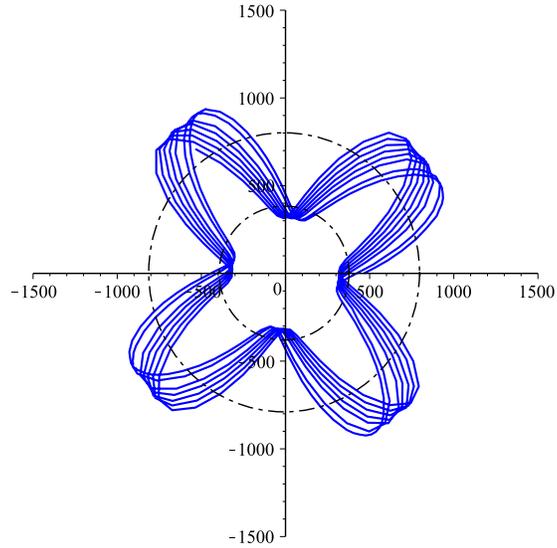}
\caption
{$E=2$, $L=0.5$, Bound orbit in region (C) for particle.}
\label{O4}
\end{figure}
\begin{figure}
\centering
\includegraphics[width=7.5cm]{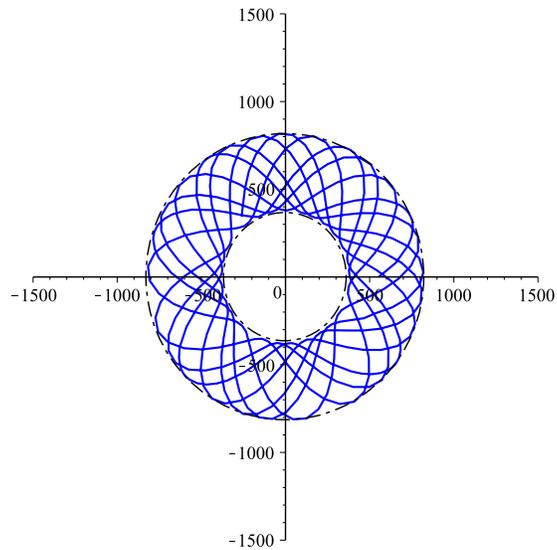}
\caption
{$E=0.5$, $L=0.5$, Bound orbit in region (D) for particle. The turning points coincide with the event horizons}
\label{O5}
\end{figure}
\begin{figure}[!ht]
\centering
\subfigure{\includegraphics[width=7.5cm]{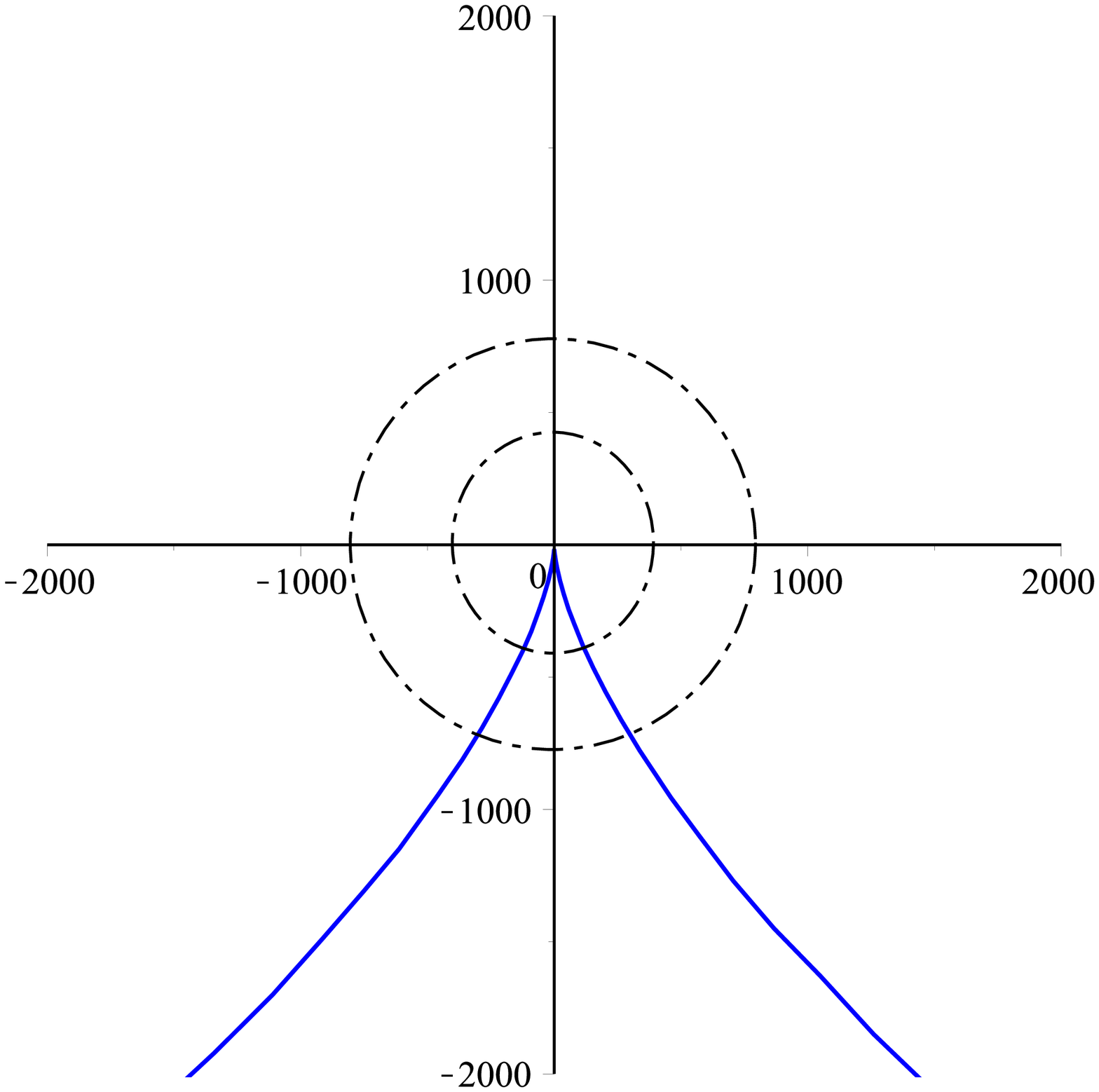}\label{O6}}

\subfigure{\includegraphics[width=7.5cm]{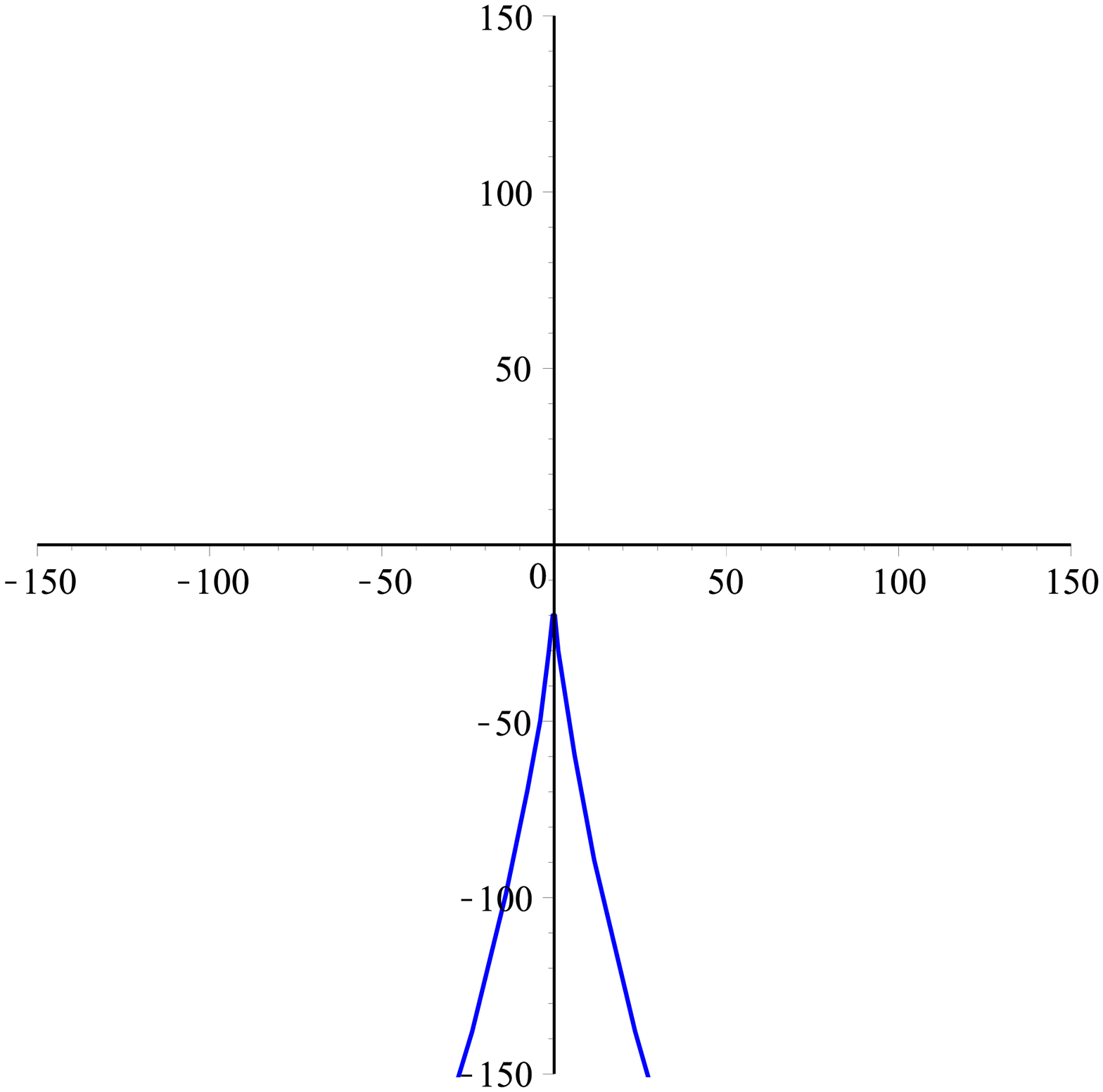}\label{O7}}
\caption
{Escape orbit in region (E) for light ray,\\ \subref{O6} $E=2$, $L=0.5$.
 \subref{O7} overview.}
\label{EOBTZ}
\end{figure}
\clearpage
\subsection{The 2+1 black hole orbits }
Analytical solution method which is introduced in section \ref{Asge} and the effective potential in section \ref{Sec5} and the number of zeroes of its polynomial, leading us to plot crossover bound orbit for the 2+1 black hole which is different from what was achieved and plotted by Cruz {\it et. al.} with the same parameters as ours. Their plot depicts a terminating bound orbit \cite{Cruz:1994ar}.
\begin{figure}[!ht]
\centering
\includegraphics[width=8cm]{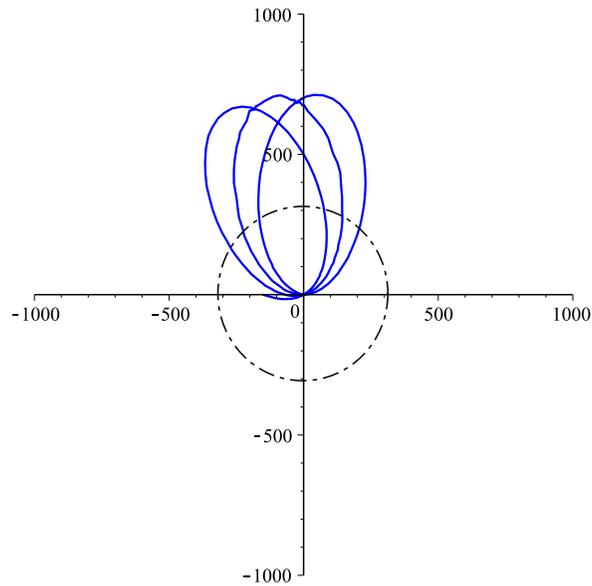}
\caption
{$E=2$, $L=2$, $J=0.5$, Crossover bound orbit in region (F) for particle.}
\label{O8}
\end{figure}
\clearpage
\section{Conclusion}
In this paper the motion of test particles and light rays in neutral Brans-Dicke Dilaton black hole in three dimensions, BTZ black hole and the 2+1 black hole were considered. We presented the analytical solution of geodesic equations for test particle motion in terms of Weierstrass elliptic function. In addition, we calculated the effective potentials. Then we showed some set of orbit types for the test particle and light rays moving on geodesics using these analytical solutions and effective potentials. As shown in figures (\ref{O1}-\ref{O8}) in region (A) we had terminating bound orbit ending in the singularity, in region (B) we had terminating orbit and in region (C and D) we showed bound orbits, also in region (E) we had escape orbit, finally we illustrated crossover bound orbit in region (F). It is noticeable that there are not any bound orbits in neutral Brans-Dicke Dilaton black hole, while there are bound orbits in 3D BTZ black hole. We should like to point out that about the main results we achieved in this paper, for a complete understanding of the physical properties of solutions of Einstein's field equations it is essential to study the orbits of test particles and light rays in rotating space-time in three dimensions and there are some advantages as we mentioned before. Meanwhile, this is important from an observational point of view, since only matter and light are observed and, as a result, can give insight into the physics of a given gravitational field. Moreover, this study was also essential from a fundamental point of view, since the motion of matter and light can be used to classify a given space-time also to decode its structure and to highlight its characteristics. Indeed, it can be shown that the space-time geometry can be constructed from the concepts of light propagation and freely falling test particles. Furthermore, we calculated the 2+1 black hole analytically and we showed the advantages of analytical method in comparison with the Cruz and {\it et. al.} method. We achieved different orbit with the same of parameter which was used by Cruz and {\it et. al.} for the 2+1 black hole and we showed a possible orbit accurately. 
To warp up, we present a useful tool to calculate the exact orbits and obtained results proved valuable in order to analyse their properties like the periastron shift of bound orbits, the light deflection of flyby orbits, the deflection angle, and the lense-Thirring effect.  
\bibliographystyle{amsplain}

\end{document}